\documentclass[preprintnumbers,amsmath,amssymb,letterpaper,nofootinbib,notitlepage]{revtex4}
\usepackage{graphicx}
\usepackage{enumerate}
\usepackage{bm}
\usepackage{slashed}

\newcommand\Eq[1]{Eq.~\ref{eq:#1}}

\usepackage{amssymb}
\usepackage{amsmath}
\usepackage{epsfig}
\newcommand{\be}{\begin{equation}}
\newcommand{\ee}{\end{equation}}
\newcommand\beq{\begin{eqnarray}}
\newcommand\eeq{\end{eqnarray}} 
\newcommand\eqn[1]{\label{eq:#1}}

\newcommand{\Tr}{{\rm Tr\,}}
\newcommand{\tr}{{\rm tr\,}}

\newcommand{\threeptwo}{${}^3P_2$ }
\newcommand{\lrnabla}{\overleftrightarrow\nabla}
\newcommand{\nn}{\nonumber}

\newcommand{\mybar}[1]%
        {\kern 0.6pt\overline{\kern -0.6pt#1\kern -0.6pt}\kern 0.6pt}
\begin{document}

\preprint{UM-DOE/ER/40762-527 }

\title{The low lying modes of triplet-condensed neutron matter and their effective theory}

\author{Paulo F. Bedaque}
\email{bedaque@umd.edu}

\author{Amy N. Nicholson}
 \email{amynn@umd.edu}
  
\affiliation{Maryland Center for Fundamental Physics, Department of Physics,
University of Maryland, College Park MD 20742-4111, USA}


 
 \begin{abstract}
The condensation of neutrons into a \threeptwo superfluid phase occurs at densities relevant for the interior of neutron stars. The triplet pairing breaks rotational symmetry spontaneously and leads to the existence of gapless modes (angulons) that are relevant for many transport coefficients and to the star's cooling properties. We derive the leading terms of the low energy effective field theory, including the leading coupling to electroweak currents, valid for a variety of possible \threeptwo phases.

\end{abstract}

\maketitle

\section{Introduction}

While understanding the phases of QCD at various densities and temperatures is at the forefront of nuclear physics research, this goal remains an elusive one due to the nonperturbative nature of QCD at all but the highest densities and temperatures. Of particular interest for the study of the cores of neutron stars is that of nuclear matter above nuclear matter density, $\rho_{nm}$, at temperatures which are low relative to the Fermi energy. For low densities interacting nucleons may be used to describe the low-temperature properties of a system, while at asymptotically high densities QCD becomes deconfined and a number of possible ground states have been proposed. At moderate densities ($\sim \rho_{nm}$) it is expected that neutrons will condense to form a superfluid state. The spontaneous breaking of any continuous symmetries by the condensate will lead to massless Goldstone bosons, which then dominate the low-energy properties of the system. 

At approximately $1.5$ times $\rho_{nm}$,  s-wave interactions, which dominate at lower densities, become repulsive and \threeptwo interactions dominate. This suggests that the order parameter for the superfluid phase in this regime will be of the form
\beq
\langle n^T\sigma_2\sigma_i\lrnabla_jn\rangle = \Delta_{ij} e^{i\alpha}\ ,
\label{eq:orderparameter}
\eeq
where $n$ are neutron field operators, the Pauli matrices act in spin space, and $\alpha$ refers to the $U(1)$ phase associated with spontaneously broken baryon number. Because the neutrons couple to form a spin-2 object, $\Delta_{ij}$ is a symmetric traceless tensor. The condensate spontaneously breaks rotational invariance, thus there will be new massless modes associated with this breaking in addition to the usual superfluid phonon\footnote{In some circumstances, like ${}^3$He, spin and orbital rotations are separate approximate symmetries and their breaking would lead to additional approximately gapless modes. In this paper we will not assume that spin and orbital rotations are separate symmetries and only the exactly gapless modes generated by the breaking of rotation symmetry ( corresponding to the diagonal group of combined spin and orbital rotations) are considered. See Sec.~\ref{sec:validity} for more discussion of this point.} (for a review concerning Goldstone bosons in systems lacking Lorentz invariance, see \cite{Brauner:2010wm}). These massless modes, referred to as angulons  \cite{Bedaque:2003wj}, have been shown to provide a mechanism for neutrino emission in neutron stars.
Recently, interest in the \threeptwo phase of neutron matter and its transport properties has been rekindled by the observation of rapid cooling of the neutron star in Cassiopeia A \cite{Heinke:2010cr}, the youngest known neutron star in the Milky Way, interpreted by two groups as evidence for triplet pairing \cite{Page:2010aw}\cite{Shternin:2010qi}\cite{Yakovlev:2010ed}\cite{Page:2012se}\cite{Page:2011yz}. A systematic understanding of the properties of \threeptwo condensed matter is necessary for sharpening this conclusion.

There are also further theoretical motivations for the calculations presented in this paper. One is that the very existence of angulons has recently been put into question \cite{Leinson:2012pn}. Also, due to the spontaneous breaking of rotational symmetry,  a large number of terms in the action are allowed by the symmetries and it does not seem possible to fix their coefficients by the usual matching procedure unless the ground state has a condensate of a special form like phase B, below.
 In fact,  reference \cite{Bedaque:2003wj} assumes the ground state to be in phase B simply to avoid the problems that the other phases raise.

While the form of the order parameter is dictated by \Eq{orderparameter}, different symmetric traceless tensors break different symmetries and there are several possible \threeptwo phases. We may choose some orthonormal frame to write down three simple symmetry breaking patterns:
\begin{subequations}\label{eq:phases}
\begin{align}
 \Delta^0&=\bar{\Delta}\left(\begin{array}{ccc}
-1/2&0&0 \\
0&-1/2&0 \\
0&0&1 \\
\end{array}\right)\quad &\text{Phase A}\eqn{phaseA}\\
\Delta^0&=\bar{\Delta}\left(\begin{array}{ccc}
e^{2i\pi/3}&0&0 \\
0&e^{-2i\pi/3}&0 \\
0&0&1 \\
\end{array}\right)\quad &\text{Phase B}\eqn{phaseB}\\
\Delta^0&=\bar{\Delta}\left(\begin{array}{ccc}
1&0&0 \\
0&-1&0 \\
0&0&0 \\
\end{array}\right)\quad &\text{Phase C}\eqn{phaseC}
 \end{align}
 \end{subequations}
In Phase A, rotational invariance is maintained in one plane, leading to two angulons associated with the breaking of rotational invariance in the remaining two planes. Phase B fully breaks rotational invariance, leading to three angulons, and was considered in \cite{Bedaque:2003wj} due to the simplicity of the effective theory for a unitary order parameter. Phase C leads to only one angulon due to the lack of a condensate in the third direction, but also contains gapless neutron modes. It is currently unclear which phase corresponds to the ground state for the relevant regions of neutron stars, and more complicated phases than the three simple ones presented here are certainly possible. 
Near the critical temperature, Ginsburg-Landau arguments can be applied and the form of the condensate is known to be a {\it real} symmetric matrix  \cite{Richardson:1972xn}. Estimating the coefficients of the Ginsburg-Landau free energy by the BCS approximation (weak coupling) one finds that phase A is favored (phase C is a close second). Strong coupling corrections to BCS reinforce this conclusion \cite{Vulovic:1984kc}. At lower temperatures the problem is more complicated, even in the BCS approximation. However, it was pointed out in \cite{Takatsuka:1992ga}\cite{Khodel:1998hn} that, when mixing between \threeptwo and ${}^3F_2$ channels can be neglected, the relative ordering between the different \threeptwo phases is independent of temperature, density and even neutron-neutron interactions. The ${}^3F_2-{}^3P_2$ mixing alters this result somewhat by lifting some degeneracies\cite{Khodel:2001yi}. In view of this uncertainty on the precise form of the condensate we will try to be as general as possible and derive the effective theory for a general {\it nodeless} phase (one where $\Delta^0$ has no zero eigenvalues). This generality can be maintained only up to some point. Final explicit expressions for the numerical values of the coefficients will be given for phase A although they can be readily obtained for any other phase using the same methods.

 Because it is not clear whether an effective theory may be derived for a general phase using the standard matching procedure, we choose instead the less elegant way of deriving the effective theory directly from a microscopic model by performing a derivative expansion to eliminate high momentum modes. In this way, both the form of the effective Lagrangian and the couplings may be determined simultaneously. The result may seem to depend strongly on the choice of the microscopic model but, in fact, most of the dependence is embedded in the value of the neutron gap ($\bar \Delta$ above). By writing the effective theory coefficients in terms of the value of the neutron gap most of the dependence on the microscopic model disappears.

\section{Microscopic model}

To derive an effective theory describing the low-energy modes of neutron matter in a \threeptwo condensed phase directly from QCD is not currently possible due to its nonperturbative nature. However, as we are only interested in low-energy properties near the Fermi surface it is sufficient to begin with a model which encapsulates the relevant properties. We choose a simple model which reproduces the leading order low-energy observables, the Fermi speed and the gap, consisting of two species (corresponding to spin states) of non-relativistic neutrons with an attractive, short range potential and a common chemical potential, 
\beq\label{eq:L}
\mathcal{L} = \psi^{\dagger}\left(i\partial_0-\epsilon(-i\nabla)\right)\psi -\frac{g^2}{4}\left(\psi^{\dagger}\sigma_i\sigma_2\lrnabla_j\psi^{*}\right)\chi_{ij}^{kl}\left(\psi^T\sigma_2\sigma_k\lrnabla_l\psi\right) \ ,
\eeq
where $\chi_{ij}^{kl} = \frac{1}{2}( \delta_{ik}\delta_{jl}+\delta_{il}\delta_{jk}-\frac{2}{3}\delta_{ij}\delta_{kl})$ is the projector onto the \threeptwo channel satisfying $\chi_{ij}^{kl} \chi_{mn}^{ij} =\chi_{mn}^{kl}$.

There are two ways of interpreting the calculation we are about to describe. One is to take $\epsilon(p) = p^2/2M-\mu$ (or its relativistic counterpart) and adjust the coupling $g^2$ so the vacuum neutron-neutron \threeptwo phase-shift is reproduced. This would make eq.~(\ref{eq:L}) a reasonable schematic model of neutron matter leading to pairing in the \threeptwo channel.
The model would correctly predict the form of the effective theory for the angulons, as well as give an estimate of the value of the low energy coefficients appearing in it. 

We will argue, however, that our calculation can be placed in a more rigorous framework, showing that it is likely capable of more quantitative predictions. Fermi liquid theory \cite{migdal1967theory}, a theory for the low lying excitations around the Fermi surface, can be cast as the effective theory obtained by integrating out neutron modes far from the Fermi surface \cite{Shankar:1994qy,Benfatto:1990kx,Polchinski:1992vn}. The explicit degrees of freedom of the Fermi liquid theory of neutron matter are quasiparticles with neutron quantum numbers whose kinetic energy is  $\epsilon(p)=v_F(p-k_F)$, where $v_F$ is the Fermi velocity and $k_F$ is the Fermi momentum.

The interactions between quasiparticles are described by interaction terms (Landau's $f_L,g_L $ parameters ), including multi-body forces whose contributions to observables are suppressed compared to those of two-body forces. The essential point is that even in systems that are strongly interacting, like neutron matter, and where perturbation theory has limited validity, it is possible to have a Fermi liquid theory description that is weakly coupled. The effect of the strong interactions is to renormalize the neutron mass and Fermi velocity, as well as change the effective interaction at the Fermi surface. If the renormalized interactions are small, the Fermi liquid description of the system is weakly coupled in the sense that the neutron quasiparticles around the Fermi surface interact weakly, despite the fact that neutrons in the bulk of the Fermi sphere are strongly interacting. In such cases the non-perturbative effects are encapsulated in the values of the Fermi velocities, effective masses, and Landau parameters. These values may then be used perturbatively in the computation of many observables.

There is evidence that the Fermi liquid effective theory of neutron matter is indeed weakly coupled \cite{Friman:2012ft}.  For instance, the fact that model calculations give pairing gaps much smaller than the Fermi energy suggests that all attractive channels are weakly coupled.  This is not surprising given that we observe that even the bare, unrenormalized nuclear phase shifts at momenta close to the Fermi surface are very modest. In any case, it will be an assumption of the present work that the interaction between quasiparticles is weak. We only keep the interaction in eq.~(\ref{eq:L}) since, as it is well known, even small attractive interactions lead to pairing and should receive special treatment (in renormalization group language the pairing interaction is marginally relevant). 

As we will see, the value of the low energy constants we derive are combinations of the density of states at the Fermi surface and geometrical factors coming from the geometry of the manifold of degenerate ground states; the quasiparticle interactions will appear only through the values of $v_F$ and $k_F$.  Other interactions not included in eq.~(\ref{eq:L}) (assuming they are indeed perturbative)  only perturbatively change the value of the low energy constants, as do higher loop effects. It would be very important to include the effect of these other interactions in a consistent manner, verify their perturbative nature and quantify its effects. Due to the difficulty involved we will leave this to another publication. Our calculation thus can be seen as one link in a chain of effective field theories connecting the phenomenology of neutron stars to ``first principles": QCD $\rightarrow$ effective theory for neutron interactions $\rightarrow$ neutron matter Fermi liquid theory $\rightarrow$ angulon effective theory$\rightarrow$ phenomenology.

Since the angulons, as Goldstone bosons, correspond to spacetime-dependent rotations of the order parameter, we will start by rewriting the theory defined by eq.~(\ref{eq:L}) in terms of the condensate. For that we introduce an auxiliary field, $\Delta_{ij}$, in the neutron pair (BCS) channel
\beq
S[\Delta,\psi]
&=&\int d^4x \left[\psi^{\dagger}\left(i \partial_0-\epsilon(-i\nabla)\right)\psi+\frac{1}{4g^2}\Delta_{ij}^{\dagger}\Delta_{ji}+\frac{\Delta_{ij}^{\dagger}}{4}\left(\psi^{T}\sigma_2\sigma_i\lrnabla_j\psi\right)-\frac{\Delta_{ji}}{4}\left(\psi^{\dagger}\sigma_i\sigma_2\lrnabla_j\psi^{*}\right)\right]\nn\cr
&=&
\int d^4x\left[
\frac{1}{4g^2}\Delta_{ij}^{\dagger}\Delta_{ji}
+\frac{1}{2}
\begin{pmatrix}
\psi^\dagger & \psi
\end{pmatrix}
\begin{pmatrix}
i \partial_0-\epsilon(-i\nabla) &   -\Delta_{ji}\sigma_i\sigma_2\nabla_j    \\
    \Delta_{ij}^{\dagger}\sigma_2\sigma_i\nabla_j          &                i \partial_0+\epsilon(-i\nabla)
\end{pmatrix}
\begin{pmatrix}
\psi \\
 \psi^*
\end{pmatrix}
\right]\ ,
\eeq
where we have dropped the projectors with the understanding that the functional integration is restricted to only the $\Delta_{ij}$ that are traceless and symmetric.  We may now perform the gaussian integration over the fermions resulting in the following action,
\beq
S[\Delta]=\int d^4x \left[\frac{1}{4g^2}\Delta_{ij}^{\dagger}\Delta_{ji} -i \Tr \ln \left( \begin{array}{cc}
i\partial_0-\epsilon(-i\nabla) & -\Delta_{ji}\sigma_i\sigma_2\nabla_j \\
\Delta_{ij}^{\dagger}\sigma_2\sigma_i\nabla_j & i\partial_0+\epsilon(-i\nabla) 
\end{array}\right) \right]\ .
\label{eq:microaction}
\eeq Up to now our calculation is exact. However, 
for a generic space-time dependent $\Delta_{ij}$ this action is complicated and highly non-local. As we are only interested in deriving a low-energy effective theory, we can obtain a useful expression if we perform a derivative expansion of $S[\Delta]$. Keeping only the leading order terms in such an expansion gives a local action in which high momentum modes have been removed. We may then parametrize the auxiliary field in terms of our effective degrees of freedom, the angulons, to find the angulon dispersion relations and interactions for a given phase.


\section{Effective theory}

Following  \cite{Fraser:1984zb,Cheyette:1985ue,Cheyette:1987fh,Chan:1985ny} we perform a derivative expansion on the logarithm in \Eq{microaction} by first separating the auxiliary field into its constant ground state plus spatial variations, $\Delta(x) \to \Delta^0+\Delta(x)$. As outlined in detail in App.~\ref{app:derivexpansion}, this leads to the following expansion for the action,
\beq\label{eq:der_exp}
S[\Delta]&=&\int d^4x \left[\frac{1}{4g^2}\Delta_{ij}^{\dagger}\Delta_{ji} -i\Tr \ln D_0^{-1}(p) \right.\cr
&-& 
 i\left.\int\frac{d^4p}{(2\pi)^4}\int_0^1dz\tr \sum_{n=0}^{\infty}(-z)^n\left[D_0(p)\sum_{m=1} \frac{\partial^{m}_{\mu}}{m!}  p_j\left[\delta D^{-1}(x)\right]_j (i\partial_{p_{\mu}})^m \right]^nD_0(p)p_k\left[\delta D^{-1}(x)\right]_k\right] \ ,
\eeq
where
\beq
D_0^{-1}(p) = 
\begin{pmatrix}
 p_0 - \epsilon(p)       &  i \Delta^0_{ji}\sigma_i\sigma_2 p_j                        \\
   - i \Delta^{0\dagger} _{ij} \sigma_2 \sigma_ip_j                                             &   p_0 +\epsilon(p)
\end{pmatrix},
\quad
\left[\delta D^{-1}(x)\right]_j = 
\begin{pmatrix}
   0  &  i \Delta_{ji} (x)\sigma_i\sigma_2                         \\
   - i \Delta^\dagger(x) _{ij} \sigma_2 \sigma_i                                             &   0
   \end{pmatrix}\ ,
\eeq 
and $\tr$ corresponds to a trace over Gorkov indices. The first two terms are the one-loop effective potential evaluated at $\Delta=\Delta^0$. The remaining terms give the space-time variation of the field $\Delta$, which describes not only the Goldstone bosons but also other, gapped degrees of freedom. Later, we will identify $\Delta = R(\alpha)\Delta^0 R^{T}(\alpha)$, where $R$ is an $SO(3)$ rotation matrix;  the Goldstone boson fields $\alpha$ are the ones parametrizing the space-time dependent rotation $R(\alpha)$. 

The leading order term in the derivative expansion contains two derivatives. This is given by the $m=2, n=1$ term in eq.~(\ref{eq:der_exp}) and leads to the following action,
\beq
S_2[\Delta] = -\frac{i}{4}\int d^4x \int\frac{d^4p}{(2\pi)^4} \tr \left[ D_0(p)\partial_{\mu} \partial_{\nu}  \delta D^{-1}(x) \partial_{p_{\mu}} \partial_{p_{\nu}} D_0(p)\delta D^{-1}(x) \right] \ ,
\label{eq:intermedaction}
\eeq
where we have dropped the constant terms associated with the vacuum energy. These terms can be minimized for constant $\Delta$ to determine which phase corresponds to the ground state. However, as discussed in the Introduction, there already exists extensive literature on this issue, including the effects of the non-zero coupling to interactions in the $^3F_2$ channel \cite{Khodel:2001yi,Takatsuka:1992ga,Khodel:1998hn}.

From \Eq{intermedaction} we disentangle the dependence on the field $\Delta$ by performing the matrix multiplication. Upon integrating by parts we find,
\beq
S_2[\Delta] &=&\int d^4x \left[\mathcal{A}_{\mu,i,j,\nu,k,l} \partial_{\mu}\Delta^{\dagger}_{ij}\partial_{\nu}\Delta^{\dagger}_{kl} \right.+  \mathcal{B}_{\mu,i,\nu,j}\left[\partial_{\mu}\Delta \cdot \partial_{\nu}\Delta^{\dagger}\right]_{ij} + \left.\mathcal{A}^{\dagger}_{\mu,i,j,\nu,k,l} \partial_{\mu}\Delta_{ji}\partial_{\nu}\Delta_{lk} \right] \ ,
\eeq
where we have used the shorthand $D_{ab} \equiv \left[D_0(p)\right]_{ab}$ ($a,b$ are Gorkov indices), and the coefficients are given by
\beq
 \mathcal{A}_{\mu,i,j,\nu,k,l}&\equiv& -\frac{i}{4}\int\frac{d^4p}{(2\pi)^4} \tr\left[\partial_{p_{\mu}}\left(D_{12}p_j\right) \sigma_2\sigma_i \partial_{p_{\nu}}\left(D_{12}p_l\right)\sigma_2\sigma_k \right], \cr
\mathcal{B}_{\mu,i,\nu,j} &\equiv& -\frac{i}{4}\int\frac{d^4p}{(2\pi)^4} \tr\left[-2\partial_{p_{\mu}}\left(D_{11}p_i\right)  \partial_{p_{\nu}}\left(D_{22}p_j\right)\right]\ .
\eeq
The derivatives of the propagator are
\beq
\partial_{p_{\mu}}D_{11}(p) &=&\frac{1}{(p_0^2-E_p^2)^2}\left[\left(-p_0^2-2 p_0 \epsilon(p)-E_p^2 \right)\delta_{\mu,0}\right.\cr
&+&\left.\left(v_F \frac{p_k}{p}((p_0+\epsilon(p))^2-p\cdot\Delta^{0\dagger}\Delta^0\cdot p)+2 (p_0+\epsilon(p)) p\cdot\Delta^{0\dagger} \Delta^0_k\right)\delta_{\mu,k}\right] \cr
\partial_{p_{\mu}}D_{12}(p) &=&\frac{1}{(p_0^2-E_p^2)^2}\left[\left( 2 i p_0 \Delta^0_{ji}  \sigma_i\sigma_2 p_j\right)\delta_{\mu,0}\right.\cr
&-& \left.i  \Delta^0_{ji}  \sigma_i\sigma_2 \left[2 \epsilon(p) v_F \frac{p_k p_j}{p} + (p_0^2-E^2_p)\delta_{jk}+2 \left(p.\Delta^{0\dagger} \Delta^0\right)_k  \right]\delta_{\mu,k}\right] \cr
\partial_{p_{\mu}}D_{22}(p) &=&\frac{1}{(p_0^2-E_p^2)^2}\left[\left(-p_0^2+2 p_0 \epsilon(p)-E_p^2\right)\delta_{\mu,0}\right.\cr
&+&\left.\left(v_F \frac{p_k}{p}(-(p_0-\epsilon(p))^2+p\cdot\Delta^{0\dagger}\Delta^0\cdot p)+2 (p_0-\epsilon(p)) p\cdot\Delta^{0\dagger} \Delta^0_k  \right)\delta_{\mu,k}\right] ,\cr
\eeq
with the definition $E_p\equiv \sqrt{\epsilon(p)^2 + p\cdot\Delta^{0\dagger}\Delta^0\cdot p}$. We find that the coefficients for the temporal derivative terms are given by the integrals
\beq
 \mathcal{A}_{0,i,j,0,k,l}&=&\Delta^0_{ca}\Delta^0_{db}(\delta_{ai}\delta_{bk}-\delta_{ab}\delta_{ik}+\delta_{ak}\delta_{ib})a_{ajbl}\cr
a_{ajbl}&\equiv&2i\int\frac{d^4p}{(2\pi)^4} \frac{p_0^2p_ap_jp_b p_l}{(p_0^2-E_p^2)^4} =  -\frac{1}{16}\int\frac{d^3p}{(2\pi)^3}\frac{p_ap_jp_bp_l}{E_p^5} \approx -\frac{Mk_F}{24\pi^2\bar{\Delta}^2}\mathcal{I}^{(2)}_{ajbl} \cr
\mathcal{B}_{0,i,0,j}&=&i\int\frac{d^4p}{(2\pi)^4} \frac{p_ip_j}{(p_0^2-E_p^2)^4}\left((p_0^2+E_p^2)^2-4p_0^2\epsilon(p)^2\right) \cr
&=& \frac{1}{8}\int\frac{d^3p}{(2\pi)^3} \frac{p_ip_j(2\epsilon(p)^2+p\cdot\Delta^0\Delta^{0\dagger}\cdot p)}{(\epsilon(p)^2+p\cdot\Delta^0\Delta^{0\dagger}\cdot p)^5}\approx \frac{Mk_F}{6\pi^2\bar{\Delta}^2}\mathcal{I}^{(1)}_{ij} \ ,
\eeq
where we have used the fact that, for small $\bar{\Delta}/v_F$, the integral is dominated by the singularity at $p=k_F$ to make the approximations, $p\approx k_F, \epsilon(p) \approx v_f(p-k_F)$. Besides the derivative expansion this is the only other approximation made up to now. Although the value of the neutron gap is a famously difficult quantity to compute, there is no question that the value of the neutron gap is below $\approx 2$ MeV and is much smaller than the Fermi energy \cite{PhysRevC.58.1921,Takatsuka:1992ga,Amundsen:1984qc,PhysRevC.78.015805}. We have also defined the remaining angular integrals as
\beq
\mathcal{I}^{(\alpha)}_{ij\cdots}(\hat{\Delta}^{0 \dagger}\hat{\Delta}^0) \equiv \int \frac{d\hat{p}}{4\pi}\frac{\hat{p}_i \hat{p}_j \cdots}{\left(\hat{p}\cdot\hat{\Delta}^{0 \dagger}\hat{\Delta}^0 \cdot \hat{p}\right)^{\alpha}}\ ,
\eeq
where $\hat{\Delta}^0 \equiv \Delta^0/\bar{\Delta}$ and $\hat p_i=p_i/p$. These integrals are functions of $\hat{\Delta}^{\dagger} \hat{\Delta}$, and depend on which phase is considered (to be more precise, they depend on the  squares  of the eigenvalues of $\hat \Delta$) so we will postpone their evaluation until the next section.

The spatial derivative terms in the Lagrangian are given by:
\beq
 \mathcal{A}_{a,i,j,b,k,l}&=&\Delta^0_{mc}\Delta^0_{nd}(\delta_{ci}\delta_{dk}-\delta_{cd}\delta_{ik}+\delta_{ck}\delta_{id})a_{abjmnl}\cr
a_{abjmnl}&\equiv& \frac{i}{2}\int\frac{d^4p}{(2\pi)^4} \frac{1}{(p_0^2-E_p^2)^4}\left[\left(p_j (2\epsilon(p)v_F\frac{p_ap_m}{p}+(p_0^2-E_p^2)\delta_{ma}+2p_m\left[\Delta^{0\dagger}\Delta^0\cdot p\right]_a)-\delta_{aj}(p_0^2-E_p^2)p_m\right)\right.\cr
&\times&\left.\left(p_l(2\epsilon(p)v_F\frac{p_bp_n}{p}+(p_0^2-E_p^2)\delta_{bn}+2p_n\left[\Delta^{0\dagger}\Delta^0\cdot p\right]_b)-\delta_{lb}(p_0^2-E_p^2)p_n\right)\right] \cr
&\approx& \frac{Mk_Fv_F^2}{24\pi^2\bar{\Delta}^2} \mathcal{I}^{(2)}_{abjmnl}\left[1 +\mathcal{O}(\bar{\Delta}^2/v_F^2)\right] 
\eeq
\beq
&&\mathcal{B}_{a,i,b,j}=i\int\frac{d^4p}{(2\pi)^4} \frac{1}{(p_0^2-E_p^2)^4}\left[v_F\frac{p_ap_i}{p}\left[(p_0+\epsilon(p))^2-4p\cdot\Delta^{\dagger}\Delta\cdot p\right]+2p_j(p_0+\epsilon(p))\left[p\cdot\Delta^{\dagger}\Delta\right]_a\right.\cr
&+&\left.\delta_{ai}(p_0+\epsilon(p))(p_0^2-E_p^2)\right]\left[v_F\frac{p_bp_j}{p}\left[-(p_0-\epsilon(p))^2-4p\cdot\Delta^{\dagger}\Delta\cdot p\right]+2p_j(p_0-\epsilon(p))\left[p\cdot\Delta^{\dagger}\Delta\right]_b+\delta_{bj}(p_0-\epsilon(p))(p_0^2-E_p^2)\right] \cr
&&\qquad \approx -\frac{Mk_fv_F^2}{6\pi^2\bar{\Delta}^2}\mathcal{I}^{(1)}_{abij}\left[1+\mathcal{O}(\bar{\Delta}^2/v_F^2)\right] +\frac{Mk_F}{\pi^2}\ln \left(\frac{\Lambda}{k_f\bar\Delta}\right) \delta_{ai}\delta_{bj}\ ,
\eeq
where $\Lambda$ is an ultraviolet cutoff of the order of the breakdown scale of the effective theory, namely, $\Lambda \approx k_f \bar\Delta$. This term is suppressed by $\sim \bar\Delta^2/v_f^2$ compared to the remaining ones and we will subsequently drop it.

By performing the index contractions, we can find the effective action up to two derivative terms in an expansion around the point $x=0$. However, the action is space-time translation invariant and its form at $x=0$ determines it at any other space-time point. As explained in  \cite{Cheyette:1987fh,Chan:1985ny,Das:1994nt}, the effective action is then given by dropping all undifferentiated $\delta D(x)$ and substituting $\Delta$ for $\Delta^0$. The final result is
that the general form for the effective theory to second order in a derivative expansion (and up to  terms of order $\mathcal{O}(\bar{\Delta}^2/v_F^2)$ and higher) is:
\beq
S_2[\Delta] &=&\frac{ M k_F}{12\pi^2\bar\Delta^2} \int d^4x\left[\mathcal{I}^{(1)}_{ij}(\hat{\Delta}^{\dagger}\hat{\Delta})\left[\partial_0\Delta\cdot\partial_0\Delta^{\dagger}\right]_{ij} - v_F^2 \mathcal{I}^{(1)}_{ijkl}(\hat{\Delta}^{\dagger}\hat{\Delta})\left[\partial_k\Delta\cdot\partial_l\Delta^{\dagger}\right]_{ij} \right.\cr
&&\qquad\qquad\qquad
+ \frac{1}{2}\mathcal{I}^{(2)}_{ijkl}(\hat{\Delta}^{\dagger}\hat{\Delta})\left(-2\left[\hat{\Delta} \cdot \partial_0\Delta^{\dagger}\right]_{ij}\left[\hat{\Delta}\cdot \partial_0\Delta^{\dagger}\right]_{kl}+\left[\partial_0\Delta^{\dagger}\cdot\partial_0\Delta^{*}\right]_{ij}\left[\hat{\Delta}\cdot\hat{\Delta}^T\right]_{kl}\right) \cr
&&\qquad\qquad\qquad
+ \left. \frac{v_F^2}{2}\mathcal{I}^{(2)}_{ijklmn}(\hat{\Delta}^{\dagger}\hat{\Delta})\left(2\left[\hat{\Delta} \cdot \partial_k\Delta^{\dagger}\right]_{ij}\left[\hat{\Delta}\cdot \partial_l\Delta^{\dagger}\right]_{mn}-\left[\partial_k\Delta^{\dagger}\cdot\partial_l\Delta^{*}\right]_{ij}\left[\hat{\Delta}\cdot\hat{\Delta}^T\right]_{mn}\right)+\mbox{h.c.}\right]\ 
\label{eq:genaction},
\eeq where $\hat\Delta \equiv \Delta/\bar\Delta$.


\section{Results for Phase A}
We will now specialize to the phase in which the eigenvalues of $\Delta^0$ are $\{-1/2,-1/2,1\}$, however, the method below can be easily carried through for any other nodeless phase. In the absence of external currents, we will see that at tree level the theory is governed by three quantities, $v_F, M$ and the ``decay constant'' $f$, which will be chosen later in order to simplify the expressions. From these three low-energy constants many physical observables may be computed, such as the specific heat (derived in Sec.~\ref{sec:kinetic}) or transport coefficients. 

We first observe that, in the case where $\Delta^0$ has two identical eigenvalues, the integrals $\mathcal{I}^{(\alpha)}_{ij\cdots}(\hat{\Delta}^{\dagger}\hat{\Delta}) $ can be written as
\beq\label{eq:Is}
\mathcal{I}^{(\alpha)}_{ij}(\hat{\Delta}^{\dagger}\hat{\Delta}) &=& A^{(\alpha)}\delta_{ij}+B^{(\alpha)}(\hat{\Delta}^{\dagger}\hat{\Delta})_{ij} \cr
\mathcal{I}^{(\alpha)}_{ijkl}(\hat{\Delta}^{\dagger}\hat{\Delta}) &=& C^{(\alpha)}\delta_{ij}\delta_{kl}+D^{(\alpha)}\delta_{ij}(\hat{\Delta}^{\dagger}\hat{\Delta})_{kl} +E^{(\alpha)}(\hat{\Delta}^{\dagger}\hat{\Delta})_{ij}(\Delta^{\dagger}\Delta)_{kl} + {\rm perm.} \cr
\mathcal{I}^{(\alpha)}_{ijklmn}(\hat{\Delta}^{\dagger}\hat{\Delta}) &=& F^{(\alpha)}\delta_{ij}\delta_{kl}\delta_{mn}+G^{(\alpha)}\delta_{ij}\delta_{kl}(\hat{\Delta}^{\dagger}\hat{\Delta})_{mn} +H^{(\alpha)}\delta_{ij}(\hat{\Delta}^{\dagger}\hat{\Delta})_{kl}(\hat{\Delta}^{\dagger}\hat{\Delta})_{mn} + {\rm perm.} \cr
&+&J^{(\alpha)}(\hat{\Delta}^{\dagger}\hat{\Delta})_{ij}(\hat{\Delta}^{\dagger}\hat{\Delta})_{kl}(\hat{\Delta}^{\dagger}\hat{\Delta})_{mn} + {\rm perm.}  \ ,
\eeq
where ``+ perm." indicates that all permutations of the indices should be included (the last term, for instance, has its $6$ indices combined in all $720$ possible ways). Numerical values for the coefficients $A^{(\alpha)},B^{(\alpha)}\cdots$ are given in the appendix. 

In phase A, rotation invariance is only partially broken, with invariance under rotation in the $(x,y)$-plane preserved. Thus, we have only two angulons, $\alpha_{1,2}$, in addition to the usual phonon. We may parametrize the field as
\beq
\Delta = e^{-i\left(\alpha_1(x)J_1+\alpha_2(x)J_2\right)/f}\Delta^0e^{i\left(\alpha_1(x)J_1+\alpha_2(x)J_2\right)/f}\ ,
\eeq
where $J_{1,2}$ correspond to the generators of infinitesimal rotations about the $x$- and $y$-axes, respectively. Here we will only consider the effective theory for the angulons, corresponding to spontaneously broken $SO(3)$ rotation symmetry. The theory for the phonon associated with breaking of $U(1)$ baryon number decouples from that of the angulons and may be treated separately. The effective theory for the phonon is much simpler and its parameters can be determined by matching as done, in the context of neutron triplet pairing, in \cite{Bedaque:2003wj}. In fact, a much more general result can be obtained by general field theoretical arguments \cite{Son:2002zn}. We will ignore the superfluid phonon from now on.
 
\subsection{\label{sec:kinetic}Kinetic terms and specific heat}

A derivative expansion of our Lagrangian in terms of the angulon fields to second order gives

\beq
S_2[\Delta] &=& \frac{1}{f^2}\frac{Mk_F}{6\pi^2}\int d^4x \left[\frac{9}{16}\left(8A^{(1)}+5B^{(1)}+80C^{(2)}+62D^{(2)}+53E^{(2)}\right)\left[(\partial_0\alpha_1)^2+(\partial_0\alpha_2)^2\right] \right.\cr
&+&v_F^2\left[-\frac{9}{64}(8(32C^{(1)}+14D^{(1)}+5E^{(1)}+288F^{(2)}+162G^{(2)}+90H^{(2)})+333J^{(2)})\left[(\partial_y\alpha_2)^2+(\partial_x\alpha_1)^2\right]\right. \cr
&-&\frac{9}{32}(8(16C^{(1)}+4D^{(1)}+E^{(1)}+96F^{(2)}+30G^{(2)}+9H^{(2)})+21J^{(2)})\left[\partial_x\alpha_1\partial_y\alpha_2+\partial_y\alpha_1\partial_x\alpha_2\right] \cr
&-& \frac{9}{8}(64C^{(1)}+58D^{(1)}+52E^{(1)}+912F^{(2)}+852G^{(2)}+801H^{(2)}+759J^{(2)})\left[(\partial_z\alpha_1)^2+(\partial_z\alpha_2)^2\right]\cr
&-&\left.\left.\frac{9}{64}(8(64C^{(1)}+22D^{(1)}+7E^{(1)}+480F^{(2)}+222G^{(2)}+108H^{(2)})+375J^{(2)})\left[(\partial_y\alpha_1)^2+(\partial_x\alpha_2)^2\right]\right]\right]\cr
&=&\int d^4x\left[\left(3+\frac{\pi}{\sqrt{3}}\right)\left[(\partial_0\alpha_1)^2+(\partial_0\alpha_2)^2\right]+v_F^2\left[\left(\frac{\pi}{9\sqrt{3}}-\frac{3}{2}\right)\left[(\partial_z\alpha_1)^2+(\partial_z\alpha_2)^2\right]\right.\right.\cr
&-&\frac{4\pi}{3\sqrt{3}}\left[(\partial_y\alpha_1)^2+(\partial_x\alpha_2)^2\right]+\left(\frac{2\pi}{9\sqrt{3}}-\frac{3}{2}\right)\left[(\partial_y\alpha_2)^2+(\partial_x\alpha_1)^2\right]\cr
&+& \left.\left.\left(\frac{3}{2}-\frac{14\pi}{9\sqrt{3}}\right)\left[\partial_x\alpha_1\partial_y\alpha_2+\partial_y\alpha_1\partial_x\alpha_2\right]\right)\right]\ ,
\label{angkinetic}
\eeq where in the second line we made the choice 
\beq
f^2 = \frac{Mk_F}{6\pi^2}.
\eeq
Note that this action is symmetric under the interchange $\{x,1\}\leftrightarrow \{y,2\}$, in accordance with our expectation of a preserved rotation symmetry in the $(x,y)$-plane. 

The condensate mixes the two angulons through the spatial derivative terms. The angulon dispersion relations may be found by diagonalizing the following matrix,
\beq
G(p) = \left(\begin{array}{cc}
a p_0^2+v_F^2( b p_z^2 + c p_y^2 + d p_x^2)  &     e v^2 p_x p_y\\
 e v^2 p_x p_y            &    a p_0^2+v_F^2(b p_z^2 + c p_x^2 + d p_y^2)
\end{array}\right) \ ,\cr
\eeq with
\beq
a &=& 3+\frac{\pi}{\sqrt{3}},\qquad b = -\frac{3}{2}+\frac{\pi}{9\sqrt{3}},\qquad c = -\frac{4\pi}{3\sqrt{3}},\cr
d &= &-\frac{3}{2}+\frac{2\pi}{9\sqrt{3}},\qquad e = \frac{3}{2}-\frac{14\pi}{9\sqrt{3}}.
\eeq
The values of $p_0$ that make the determinant of $G(p)$ vanish correspond to the poles of the angulon propagator and define their dispersion relations. As expected, the energies are proportional to the the Fermi velocity $v_F$ times spatial momenta, but there is no expectation that the velocity of the angulons will be independent of the direction. In fact,  in the particular case where the propagation is along the axes $x,y$, and $z$ the corresponding velocities (for the two modes $1$ and $2$) are
\beq
v^{(1)}_{x,y}&=& \frac{v_F}{3} \sqrt{\frac{117}{18+2\sqrt{3}\pi}-2} \approx 0.477 v_F,\cr
v^{(2)}_{x,y}&=& 2v_F\sqrt{\frac{\pi}{9\sqrt{3}+3\pi}} \approx 0.709 v_F,\cr
v^{(1,2)}_z &=& \frac{v_F}{3} \sqrt{\frac{99}{18+2\sqrt{3}\pi}-1} \approx 0.519 v_F.
\eeq
The dispersion relations for modes moving in a general direction are
\beq
p^{(1)}_0 = \frac{\sqrt{27\sqrt{3}|p|^2-2\pi[2(p_x^2+p_y^2)+p_z^2]}}{3\sqrt{2(3\sqrt{3}+\pi)}}v_F \cr
p^{(2)}_0 = \frac{\sqrt{24\pi(p_x^2+p_y^2)+27\sqrt{3}p_z^2-2\pi p_z^2}}{3\sqrt{2(3\sqrt{3}+\pi)}}v_F \ .
\eeq
The angulon modes have linear dispersion relations at small momenta, which may be used to compute the angulon contribution to the low temperature specific heat. In fact, it is given by
\beq
c_v &=& \sum_{a=1,2} \frac{d}{dT} \int \frac{d^3p}{(2\pi)^3}\frac{\epsilon_a(p)}{e^{\epsilon_a(p)/T}-1}\cr
&\approx& 16.16 \frac{T^3}{v_F^3} = 1.44\times 10^{-13}\left(\frac{T/K}{v_F/c}\right)^3 \frac{\mbox{erg}}{K\mbox{cm}^3},
\eeq where $\epsilon_{1,2}$ is the energy of the two uncoupled angulons. The dependence $c_v \sim T^3/v_F^3$ follows from dimensional analysis; the numerical coefficient comes from a numerical integration. For temperatures well below the condensation temperature for neutrons in neutron stars the specific heat due to electrons dominates \cite{Yakovlev:1999sk}, and the angulon contribution is a few orders of magnitude smaller.

\subsection{Angulon interactions}

The leading order effective action shown in \Eq{genaction} also describes interactions between angulons. Since the gapless modes, like the angulons, dominate transport processes at small temperatures, their interaction is relevant for the calculations of these quantities.  The somewhat tedious process of expanding the action to quartic order in the angulon fields leads to 
\beq
S_4[\Delta] &=& \frac{1}{f^2}\int d^4x \left[\left(3+\frac{\pi }{\sqrt{3}}\right) \left(\text{$\alpha_2$}^2(\text{$\partial_0$$\alpha_1$})^2+\text{$\alpha_1$}^2 (\text{$\partial_0$$\alpha_2$})^2\right)+\left(12+\frac{4 \pi }{\sqrt{3}}\right) \left(\text{$\alpha_1$}^2 (\text{$\partial_0$$\alpha_1$})^2+\text{$\alpha_2$}^2 (\text{$\partial_0$$\alpha_2$})^2\right)\right.\cr
&+&\left(18+2 \sqrt{3} \pi \right) \text{$\alpha_1$}\text{$\alpha_2$} \text{$\partial_0$$\alpha_1$} \text{$\partial_0$$\alpha_2$}+v_F^2\left[\left(\frac{\pi  }{9 \sqrt{3}}-\frac{3}{2}\right) \left(\text{$\alpha_2$}^2 (\text{$\partial_x$$\alpha_1$})^2+\text{$\alpha_1$}^2 (\text{$\partial_y$$\alpha_2$})^2\right)\right.\cr
&+&\left(\frac{8 \pi }{9 \sqrt{3}}-6 \right) \left(\text{$\alpha_1$}^2(\text{$\partial_x$$\alpha_1$})^2+\text{$\alpha_2$}^2 (\text{$\partial_y$$\alpha_2$})^2\right)-\frac{4\pi}{3\sqrt{3}}   \left(\text{$\alpha_1$}^2 (\text{$\partial_x$$\alpha_2$})^2+\text{$\alpha_2$}^2 (\text{$\partial_y$$\alpha_1$})^2\right)\cr
&+&\left(3 -\frac{28 \pi  }{9\sqrt{3}}\right) \left(\text{$\alpha_1$}^2 \text{$\partial_x$$\alpha_1$}\text{$\partial_y$$\alpha_2$}+\text{$\alpha_2$}^2 \text{$\partial_x$$\alpha_1$}\text{$\partial_y$$\alpha_2$}+\text{$\alpha_1$}^2 \text{$\partial_x$$\alpha_2$}\text{$\partial_y$$\alpha_1$}+\text{$\alpha_2$}^2 \text{$\partial_x$$\alpha_2$}\text{$\partial_y$$\alpha_1$}\right)\cr
&+&\left(3 -\frac{10 \pi }{3 \sqrt{3}}\right) (\text{$\alpha_1$} \text{$\alpha_2$}\text{$\partial_x$$\alpha_1$} \text{$\partial_y$$\alpha_1$}+\text{$\alpha_1$} \text{$\alpha_2$} \text{$\partial_x$$\alpha_2$} \text{$\partial_y$$\alpha_2$})-\left(\frac{16 \pi }{9 \sqrt{3}}+6 \right) (\text{$\alpha_1$}\text{$\alpha_2$} \text{$\partial_x$$\alpha_1$} \text{$\partial_x$$\alpha_2$}+\text{$\alpha_1$} \text{$\alpha_2$} \text{$\partial_y$$\alpha_1$} \text{$\partial_y$$\alpha_2$})\cr
&-&\left(\frac{35 \pi  }{9 \sqrt{3}}+\frac{3}{2}\right) \left(\text{$\alpha_2$}^2 (\text{$\partial_x$$\alpha_2$})^2+\text{$\alpha_1$}^2 \text{$\partial_y$$\alpha_1$}^2\right)+\left(\frac{2 \pi }{9 \sqrt{3}}-\frac{3 }{2}\right)\left(\text{$\alpha_2$}^2 (\text{$\partial_z$$\alpha_1$})^2+\text{$\alpha_1$}^2(\text{$\partial_z$$\alpha_2$})^2\right)\cr
&-&\left.\left.\left(\frac{\pi }{\sqrt{3}}+\frac{9 }{2}\right)\left(\text{$\alpha_1$}^2 (\text{$\partial_z$$\alpha_1$})^2+\text{$\alpha_2$}^2(\text{$\partial_z$$\alpha_2$})^2\right)-\left(\frac{22 \pi }{9 \sqrt{3}}+6 \right)\text{$\alpha_1$} \text{$\alpha_2$}\text{$\partial_z$$\alpha_1$} \text{$\partial_z$$\alpha_2$} \right]\right]\ .
\eeq


\subsection{Weak interactions}

Angulons couple to electroweak currents. Since they are not electrically charged, at leading order in the Fermi constant $G_F$ the only possible coupling is with the neutral current mediated by the Z boson. In this section we derive this coupling.

We begin by adding the following interaction terms to the microscopic Lagrangian,
\beq
\mathcal{L}_W = C_V Z_0^0 \psi^{\dagger}\psi+C_A Z_i^0 \psi^{\dagger}\sigma_i \psi \ ,
\eeq
where $Z_0^0$, $Z_i^0$ are the temporal and spatial components, respectively, of the $Z^0$ boson, and the couplings are given by
\beq
C_{V,A}^2 = \tilde{C}_{V,A}^2\frac{G_FM_Z^2}{2\sqrt{2}}\ , 
\eeq
where $\tilde{C}_V = -1$ by vector current conservation, and $\tilde{C}_A\sim 1.1\pm 0.15$ \cite{Savage:1996zd} is given by the sum of the nucleon isovector axial coupling, $g_A$, and the matrix element of the strange axial-current in the proton, $\Delta s$. Here we choose the vacuum form of the interactions as little is known about their renormalization when modes far from the Fermi surface are removed.

The action for the angulons including the weak vertex is
\beq
S[\Delta] &=& -i\int d^4x \Tr \ln [D_0^{-1}+\delta D^{-1}+C_A Z_m^0 \Sigma_m] \cr
&\approx & -i\int d^4x \Tr \left(\ln [D_0^{-1}+\delta D^{-1}]+(D_0^{-1}+\delta D^{-1})^{-1}C_A Z_m^0 \Sigma_m\right)
\eeq
where 
\beq
\Sigma_m \equiv \left( \begin{array}{cc}
\sigma_m & 0  \\
0 & \sigma_2\sigma_m\sigma_2 \\
\end{array} \right)\ ,
\eeq
and we have taken only the leading order in a weak coupling expansion. We may now perform a derivative expansion of the propagator using the method outlined in App.~\ref{app:derivexpansion},
\beq
(D_0^{-1}+\delta D^{-1})^{-1} = \int \frac{d^4p}{(2\pi)^4} \tr \left(\sum_n\left[D_0(p)\sum_{m=1}\frac{\partial_{\mu}^{m}
}{m!}p_j[\delta D^{-1}(x)]_j(i\partial_{p_{\mu}})^m\right]^nD_0(p)\right)\ .
\eeq

The leading order term in the derivative expansion of the weak interaction contribution to the Lagrangian is given by $m=n=1$. The only non-zero terms are
\beq
\mathcal{L}_W[\Delta]&=&C_AZ_m^0\int \frac{d^4p}{(2\pi)^4} \tr \left[D_0(p)\partial_0p_j[\delta D^{-1}(x)]_j\partial_{p_0}D_0(p)\Sigma_m\right] \cr
&=&iC_AZ_m^0\int \frac{d^4p}{(2\pi)^4}p_j \tr[D_{12}\partial_0\delta\Delta^{\dagger}_{ij}\sigma_2\sigma_i\partial_{p_0}D_{11}\sigma_m- D_{11}\partial_0\delta\Delta_{ji}\sigma_i\sigma_2\partial_{p_0}D_{21}\sigma_m\cr
&+& D_{22}\partial_0\delta\Delta^{\dagger}_{ij}\sigma_2\sigma_i\partial_{p_0}D_{12}\sigma_2\sigma_m\sigma_2- D_{21}\partial_0\delta\Delta_{ji}\sigma_i\sigma_2\partial_{p_0}D_{22}\sigma_2\sigma_m\sigma_2]\cr
&=& C_AZ_m^0\partial_0\delta\Delta^{\dagger}_{ij}\Delta_{kl}\int \frac{d^4p}{(2\pi)^4}p_j p_k\tr\left[\frac{-1}{p_0^2-E_p^2}\sigma_l\sigma_2\sigma_2\sigma_i\frac{-p_0^2-2p_0\epsilon_p-E_p^2}{(p_0^2-E_p^2)^2}\sigma_m\right.\cr
&+&\left.\frac{p_0-\epsilon_p}{p_0^2-E_p^2}\sigma_2\sigma_i\frac{2p_0\sigma_l\sigma_2}{(p_0^2-E_p^2)^2}\sigma_2\sigma_m\sigma_2\right]+ \mbox{h.c.}\cr
&=&-2i\epsilon_{lim}C_AZ_m^0\partial_0\delta\Delta^{\dagger}_{ij}\Delta_{kl}\int \frac{d^4p}{(2\pi)^4}p_j p_k\frac{-p_0^2+4p_0\epsilon_p+E_p^2}{(p_0^2-E_p^2)^3}+ \mbox{h.c.}\cr
&=&\frac{1}{2}\epsilon_{lim}C_AZ_m^0\partial_0\delta\Delta^{\dagger}_{ij}\Delta_{kl}\int \frac{d^3p}{(2\pi)^3}\frac{p_j p_k}{E_p^3}+ \mbox{h.c.}\cr
&\approx &\epsilon_{lim}C_AZ_m^0\partial_0\delta\Delta^{\dagger}_{ij}\Delta_{kl} \frac{Mk_F}{2\pi^2|\bar{\Delta}|^2}\mathcal{I}_{jk}^{(1)} + \mbox{h.c.}\cr
&\rightarrow& 3f^2\epsilon_{lim}C_AZ_m^0\partial_0\Delta^{\dagger}_{ij}\Delta_{kl} \mathcal{I}_{jk}^{(1)} + \mbox{h.c.},
\eeq where in the last step we used the translation invariance of the effective lagrangian, as explained previously when deriving the strong interactions. Expanding this in terms of the angulon fields $\alpha$ gives the leading contribution to the action from the angulon-neutral current vertex

\beq
S_W[\Delta] &=&C_A \int d^4x \left[\frac{27}{8}(4A^{(1)}+3B^{(1)})f (Z_2^0 \partial_0 \alpha_2 - Z_1^0 \partial_0 \alpha_1) + \frac{27}{8}(4A^{(1)}+3B^{(1)})Z_3^0 (\alpha_2 \partial_0 \alpha_1 - \alpha_1 \partial_0 \alpha_2)+\cdots\right] \cr
&=&C_A \int d^4x \left[9f(Z_2^0 \partial_0 \alpha_2 - Z_1^0 \partial_0 \alpha_1) +9Z_3^0 (\alpha_2 \partial_0 \alpha_1 - \alpha_1 \partial_0 \alpha_2)+ \cdots\right]\ .
\eeq

\subsection{\label{sec:validity}Higher orders and regime of validity}

The effective action shown in eq.~(\ref{eq:genaction}) is, to the extent that terms with more derivatives may be neglected, equivalent to our starting point (eq.~(\ref{eq:L})). We will argue now that is legitimate to use eq.~(\ref{eq:genaction}) at tree level to capture the low energy dynamics of the system. The crucial observation is that the effective action shown in eq.~(\ref{eq:genaction}) shares certain properties with other low energy effective actions, such as those describing low energy pions in zero-density/temperature QCD or magnons in an anti-ferromagnet, and power counting of diagrams in such theories is well understood. The contributions of bosonic fluctuations (angulons) are described by loops. A generic loop diagram will have an even number of powers of $f$ in the denominator coming from the vertices in eq.~(\ref{eq:genaction}). By dimensional arguments, these powers of $f$ must be offset by powers of the external momentum $Q$ so that loops are suppressed by powers of $(Q/f)^2$ compared to tree level and are therefore suppressed at low energies \footnote{The detailed accounting of powers of $Q/f$ is similar to the one in Chiral Perturbation Theory and is discussed at length in \cite{Weinberg:1979ys}. }.  

This is not to say that the range of validity of the angulon effective theory is $Q < f$. We have neglected in our derivation contributions suppressed by $\bar\Delta/v_F$ that may be larger than corrections of order $(Q/f)^2$. For astrophysical applications, where solid estimates are more necessary than precise calculations, those corrections are of limited interest.

The generic form of the effective theory we obtained could have been foreseen. It is indeed the most general action involving a symmetric tensor field $\Delta$ obeying rotation symmetry up to two derivatives. Higher powers of $\Delta$, not appearing in eq.~(\ref{eq:genaction}), are not independent since $\Delta$  has only two independent eigenvalues. The only terms not appearing in eq.~(\ref{eq:genaction}) are the ones with repeated powers of $\Delta^2$, as opposed to $\Delta^\dagger \Delta$. They are excluded because our starting point eq.~(\ref{eq:L}) has, in addition to rotational symmetry
\beq
\Delta \rightarrow R \Delta R^T,
\eeq where $R$ is a rotation matrix, the enhanced symmetry
\beq\label{eq:enhanced}
\Delta \rightarrow R_S \Delta R_L^T,
\eeq where $R_S, R_L$ are independent spin and orbital rotations. Nuclear forces do not have this enhanced symmetry (among neutrons due mostly to the existence of spin-orbit forces). As with any of the other interactions  left out of eq.~(\ref{eq:L}) (assuming they do not affect the phase of the theory) and as discussed above, they are expected to contribute to the low energy constants of the angulon theory we have calculated by a perturbatively small amount, even though they may contribute significantly to the renormalization of $k_F$ and $ v_F$. Notice that for this to be true the renormalized spin-orbit forces at the Fermi surface do not have to be smaller than the interaction in eq.~(\ref{eq:L}); it is enough that their contribution to the energy be smaller than the neutron kinetic energy, as the low energy coefficients we computed are independent of the coupling $g$.  This observation also indicates that terms like
\beq
\mathcal{L} \sim Tr(\Delta),
\eeq forbidden only by the enhanced symmetry should be small. These terms give a mass to the approximate Goldstone modes related to the enhanced symmetry eq.~(\ref{eq:enhanced}). We ignored these massive modes here entirely as they are irrelevant at small enough energies and observe that the smallness of their mass is behind some of the near degeneracies of \threeptwo states found in model calculations. Further work is needed to evaluate the approximate Goldstone masses. If it turns out that their masses are smaller than $\sim\bar\Delta/v_F$ they will set the scale of breakdown of the angulon effective theory.

\section{Summary}

We have derived a low-energy effective theory describing the Goldstone bosons associated with broken rotational symmetry in a $^3P_2$ condensed neutron superfluid (angulons). Because transport properties are dominated by the low lying excitation modes, this theory provides a link connecting the theory of nuclear forces to many quantities of interest in neutron star phenomenology. Since there is still controversy as to which of the many \threeptwo phases are realized in nature we have tried to keep our calculation as general as possible. Ultimately, however, the numerical value of the coefficients of the effective action do depend on the particular \threeptwo phase and we give explicit values for the ``phase A" as defined in 
\Eq{phases}. This effective theory is valid for angulon energies below the energy scale $\sim 2 k_f \bar\Delta$ (or the mass of the approximate Goldstone bosons related to independent spin and orbital rotations, whichever is smaller) where other degrees of freedom, like unpaired neutrons, appear. 

A simple application  of the effective theory, the calculation of the angulon contribution to the specific heat, was discussed.
We also considered the coupling of angulons to neutral currents, since quantities like neutrino opacity and emission rates depend on this coupling, and gave an explicit form for the angulon-angulon-Z vertex.

A series of improvements and extensions to the effective theory discussed here are desirable. For applications to neutron stars, we should consider the presence of both protons and neutrons. The protons are superconducting and lead only to another gapped mode but they are important in mediating the interaction between angulons (with whom they interact through strong forces) and the gapless electron (with whom they  interact electromagnetically). In our microscopic action for neutrons we have included only the dominant forces leading to \threeptwo pairing. While expected to be repulsive and weaker, neutron interactions in other channels can have an influence on the angulon effective theory. It would be very desirable to quantify this effect. We have not given much attention to the gapped modes corresponding to a change in the eigenvalues of $\Delta$. Although their importance is exponentially suppressed at small temperatures they can be numerically important at temperatures of relevance to some stages of neutron star evolution. Our method of deriving the effective theory by performing a derivative expansion on a microscopic theory allows us to address this question and we plan to come back to it in a future publication. Finally, the energy difference between different \threeptwo phases is small. In particular, the condensation energy of phase C in \Eq{phases} is only a few percent above that for phase A. This restricts the validity of the effective theory somewhat and it would be important to quantify the importance of the other nearby minima to the low energy physics of the system.

\section{Appendix: The derivative expansion}
\label{app:derivexpansion}

In order to set up the derivative expansion of
\beq
{\rm Tr}\ln D^{-1}(i\partial, x) = {\rm Tr}\ln [\underbrace{D_0^{-1}(i\partial)}_{D^{-1}(i\partial, 0)} + \delta D^{-1}(i\partial, x) ]
\eeq we first use the relation

\beq
{\rm Tr} \ln(A+B) &=& Tr\ln A + {\rm Tr} \ln (1+A^{-1}B) = Tr\ln A + {\rm Tr} \ln (1+B A^{-1})\nn\\
&=&
{\rm Tr}\ln A  + \sum_{n=0}^\infty
{\rm Tr}\frac{1}{n+1} (A^{-1}B)^{n+1}\nn\\
&=&
{\rm Tr}\ln A  + {\rm Tr}\int_0^1 dz\ (1+zA^{-1}B )^{-1}A^{-1}B\nn\\
&=&
{\rm Tr}\ln A  +{\rm Tr} \int_0^1 dz\ (A+zB )^{-1}B
\eeq to find
\beq
 {\rm Tr} \ln D_0^{-1}+
\int_0^1 dz\ {\rm Tr} \frac{1}{D_0^{-1}+z \delta D^{-1}}  \delta D^{-1}.
\label{eq:lnexpand}\eeq The second term contains the dependence on the space-time variation of $\Delta$. This term is complicated to compute because it contains $\partial$ and $x$ which do not commute. One trick to deal with this is to substitute in the integrand $\partial\rightarrow ip, x\rightarrow x+i\partial_p$ and integrate over $p$ \cite{Chan:1985ny} . To do this, we first need to find the inverse of the operator $D_0^{-1}+z\delta D^{-1}$ in terms of $p,\partial_p$, so we look at
\beq
 \left[D_0^{-1}+z\delta D^{-1}\right]^{-1} \equiv G(x,y) = \int\frac{d^4p}{(2\pi)^4} e^{ip\cdot y} G(p,i\partial_p) e^{-ip \cdot x} \ .
\eeq
Using
\beq
 \left(D_0^{-1}+z\delta D^{-1}\right) G(x,y) = \delta^4(x-y) = \int\frac{d^4p}{(2\pi)^4} e^{ip\cdot y}  \left(D_0^{-1}+z\delta D^{-1}\right) G(p,i\partial_p) e^{-ip \cdot x} \ ,
\eeq
we find
\beq
G(p,i\partial_p) = \left[D_0^{-1}(p)+z p_j \left[\delta D^{-1}(i\partial_p)\right]_j\right]^{-1} \ ,
\eeq
where
\beq
D_0^{-1}(p) &=& \left(\begin{array}{cc}
p_0-\epsilon(p) & ip_j\Delta^0_{ji}\sigma_i\sigma_2 \\
-ip_j\Delta^{0\dagger}_{ij}(\sigma_2\sigma_i & p_0+\epsilon(p) \\
\end{array} \right) \ , \cr
\left[\delta D^{-1}(i\partial_p)\right]_j &=& \left(\begin{array}{cc}
0 & i\Delta_{ji}(i\partial_p)\sigma_i\sigma_2\\
-i\Delta^{\dagger}_{ij}(i\partial_p)\sigma_2\sigma_i & 0 \\
\end{array} \right) \ .
\eeq
We may now expand the second term in \Eq{lnexpand} as
\beq
&&\int\frac{d^4p}{(2\pi)^4}\int_0^1dz\tr \left[D_0^{-1}(p)+z p_j \left[\delta D^{-1}(i\partial_p+x)\right]_j \right]^{-1}p_k\left[\delta D^{-1}(x)\right]_k \cr
&=& \int\frac{d^4p}{(2\pi)^4}\int_0^1dz\tr \left[D_0^{-1}(p)+z p_j \left[\delta D^{-1}(x)\right]_j + z p_j \sum_{m=1} \frac{(i\partial_{p_{\mu}})^m}{m!}\partial^m_{\mu} \left[\delta D^{-1}(x)\right]_j \right]^{-1}p_k\left[\delta D^{-1}(x)\right]_k\cr
&=& \int\frac{d^4p}{(2\pi)^4}\int_0^1dz\tr \left[1 + z p_j\left(D_0^{-1}(p) + z p_k\left[\delta D^{-1}(x)\right]_k\right)^{-1}\sum_{m=1} \frac{(i\partial_{p_{\mu}})^m}{m!}\partial^m_{\mu} \left[\delta D^{-1}(x)\right]_j \right]^{-1}\cr
&\times&\left[D_0^{-1}(p)+z p_j \left[\delta D^{-1}(x)\right]_j\right]^{-1}p_k\left[\delta D^{-1}(x)\right]_k \cr
&=& \int\frac{d^4p}{(2\pi)^4}\int_0^1dz\tr \sum_{n=0}^{\infty}(-z)^n\left[D_0(p)\sum_{m=1} \frac{\partial^m_{\mu}}{m!}  p_j\left[\delta D^{-1}(x)\right]_j (i\partial_{p_{\mu}})^m \right]^nD_0(p)p_k\left[\delta D^{-1}(x)\right]_k \ ,
\eeq


\section{Appendix: Coefficients for the integrals $\mathcal{I}^{(\alpha)}_{ij\cdots}$}
\label{app:coeff}
Here we show how to compute the numerical coefficients $A^{(\alpha)}, B^{(\alpha)}, \cdots$  appearing in \Eq{Is}. Consider, for instance, $\mathcal{I}^{(1)}_{ij}$. We first multiply the upper equation in \Eq{Is} by $(\hat\Delta^\dagger\hat\Delta)_{ij}$ and $\delta_{ij}$ to obtain

\beq
\underbrace{   \int \frac{d\hat p}{4\pi}\frac{1}{\hat p.(\hat\Delta^\dagger\hat\Delta).\hat p}   }_{\frac{4\pi}{3\sqrt{3}}} &=&
 3 A^{(1)}+\underbrace{   {\rm tr }(\hat\Delta^\dagger\hat\Delta)}_{3/2}     \  B^{(1)} \nn\\
\underbrace{  \int \frac{d\hat p}{4\pi}  }_{1}&=&
\underbrace{  {\rm tr }(\hat\Delta^\dagger\hat\Delta)}_{3/2}   A^{(1)} +\underbrace{  {\rm tr }(\hat\Delta^\dagger\hat\Delta)^2}_{9/8} B^{(1)}.
\eeq Solving this system of equations we find
\beq
A^{(1)}&=& \frac{4}{3}\left(\frac{\pi}{\sqrt{3}}-1\right), \qquad B^{(1)}=\frac{8}{3}-\frac{16\pi}{9\sqrt{3}}
\eeq  The same method can be easily implemented in computer algebra packages and we find

\beq
C^{(1)}&=& -\frac{4}{27}+\frac{145\pi}{1458\sqrt{3}},\qquad D^{(1)}= \frac{14}{27}-\frac{220\pi}{729\sqrt{3}}, \qquad E^{(1)}=-\frac{10}{27}+\frac{152\pi}{729\sqrt{3}}\cr
C^{(2)}&=&\frac{11}{36}-\frac{25\pi}{243\sqrt{3}}, \qquad D^{(2)}=-\frac{10}{9}+\frac{128\pi}{243\sqrt{3}}, \qquad E^{(2)}=\frac{8}{9}-\frac{112\pi}{243\sqrt{3}}\cr
F^{(2)}&=&\frac{43}{1080}-\frac{263\pi}{13122\sqrt{3}}, \qquad G^{(2)}=-\frac{59}{270}+\frac{256\pi}{2187\sqrt{3}}, \qquad H^{(2)}=\frac{16}{45}-\frac{424\pi}{2187\sqrt{3}}, \qquad J^{(2)}=-\frac{8}{45}+\frac{640\pi}{6561\sqrt{3}} .\cr
\eeq 

\begin{acknowledgments}
This work was supported in part by U.S.\ DOE grant No.\ DE-FG02-93ER-40762.
\end{acknowledgments}
\bibliography{3P2ref}
\end{document}